\documentclass[12pt]{article}
\usepackage[numbers,sort&compress]{natbib}
\usepackage{graphicx}
\usepackage{amsmath,graphicx,amssymb,subfigure}
\usepackage{amssymb}
\usepackage{amsmath}
\usepackage{enumerate}
\usepackage{amsfonts}
\usepackage[all]{xy}
\usepackage[section]{placeins}

\usepackage[colorlinks=true,linktocpage=true,linkcolor=blue,citecolor=blue]{hyperref}
\newcommand{\mt}[1]{\textrm{\tiny #1}}
\newcommand{\bea}{\begin{eqnarray}}
\newcommand{\eea}{\end{eqnarray}}
\newcommand{\rh}{r_\mt{H}}

\newcommand{\la}{\lambda}
\newcommand{\de}{\delta}

\begin{document}

\bibliographystyle{hieeetr}

\pagestyle{plain}
\setcounter{page}{1}

\begin{titlepage}

\begin{center}

\vskip 60mm

{\Large {\bf  Holographic conductivity from Einstein-Maxwell-Dilaton in Gauss-Bonnet gravity   and Entropy Function }}

\vskip 1 cm

{ {\bf Yan-Tao Hao$^1$, Li-Qing Fang$^2$, Long Cheng$^1$}}\\

\vskip .8cm

{\it $^1$ Department of Physics, Zhejiang Sci-Tech University, \\ Hangzhou 310018,  China\\
     $^2$ School of Measuring and Optical Engineering, Nanchang Hangkong University, \\Nanchang 330063, China}

\medskip

\vspace{5mm}
\vspace{5mm}

\begin{abstract}
In this paper we consider the  holographic DC and Hall conductivity in Einstein-Maxwell-Dilaton in Gauss-Bonnet gravity with momentum  dissipation. We analytically derived the DC conductivity and Hall conductivity
from the black horizon data, and found that the conductivities are independent on the Gauss-Bonnet coupling. We also used the entropy function formalism to get the conductivities in terms of the
charge of the black hole, even without knowing the explicit  black hole solutions.
\end{abstract}

\end{center}
 \noindent
\end{titlepage}

\section{Introdution}
The AdS/CFT correspondence gives a simple way to study strongly correlated systems \cite{ Hartnoll,McGreevy,Herzog:2009}. In the framework of AdS/CFT, one can relates the weakly coupled classical gravity theories to the strongly coupled quantum many-body systems. So it provides  a new approach to calculating observables in these strongly coupled  many-body system.  In particular, one can perturb the electric field and thermal gradients around black hole background to calculate the transport coefficients such as conductivities in the strongly correlated systems on the boundary.

In the past few years, there are a lot of work on  calculating the  conductivities based on several holographic models. One conventional approach is that  take the  perturbation on the boundary by a time-dependent field with frequency $\omega$ to get the optical conductivities,  and then take the zero frequency limit to get the DC conductivities. However, because of the translation invariance of the gravitational theories, there is the divergence in obtaining the DC conductivities. To avoid this unphysical diverge conductivities, one can introduce  the linear axion field in the gravitational background to brake the spatial translation invariance but remain the homogeneous \cite{withers}.  In this gravitational model, the presence of the linear axion field results the momentum dissipation in the dual boundary theory. This model has been widely used in applied holographic calculations, see the reviews \cite{Matteo} and therein.

Recently, one novel approach to calculate DC conductivities was proposed in \cite{Donos2014,Donos2014a}. Based on the idea of the membrane paradigm,  this method provides an effective way to derive the DC  transports from the  data of black horizon. So once given a charged black hole solution in AdS space, one can calculate the DC electrical conductivities in terms of  matter fields on the horizon. Furthermore, in \cite{Donos2015, Donos2015a}, the authors consider the dyonic black holes and get the Hall conductivities. Inspired by these work, various models have been discussed in \cite{Keun-Young,Yi Ling,GXH14,GXH15,GXH17,kuang1,Pope}. 
On the other hand,  for a specific class of black hole, there is an interesting property named "attractor mechanism" \cite{Sen1}. It states that  the magnitude of a matter field obtains some values on the black hole horizon, which are dependent only upon the electric and magnetic charges of the black hole. In other words, the configuration of a black hole near horizon depends only on the electric and magnetic charges of the black hole and not on the asymptotic values of other matter fields. Thus it is very natural to think about whether one can use the attractor mechanism to calculate the DC conductivities without knowing the full black hole solutions. Following the spirit of the above discussions, in the work \cite{Erd}, the authors employ the Sen's entropy function method to get the horizon data and conductivities for Einstein-Maxwell-Dilaton theory. In addition,  the Einstein-Maxwell theory with topological Maxwell term was discussed in \cite{Nas1,Nas2}. 

In this paper, following the approach \cite {Donos2014a, Donos2015} we will generalize   these discussions to more generic case. So we are going to study the  Einstein-Maxwell-Dilaton-Axion theory with Gauss-Bonnet term and find the DC and Hall conductivities  analytically. The attractor mechanism and the Sen' entropy function in  Gauss-Bonnet gravity will be performed  in calculations.   One of the motivations is that when we consider the semi-classical effect, the Gauss-Bonnet term appears as the curvature stringy correction. On the other hand, what we consider in this paper is Gauss-Bonnet gravity coupled with a nontrivial dilaton field and potential. Unlike the case of \cite{GXH14}, there is no exact black hole solution exists, so it is impossible to get the DC conductivities from horizon data directly.  We can use the entropy function and solve the attractor equations to express the conductivities at zero temperature.

This paper is organised as follows. In section 2, we introduce the Einstein-Maxwell-Dilaton-Axion theory in Gauss-Bonnet gravity. In section 3, we will calculate the DC transport coefficients such as electrical conductivity, thermal conductivity and thermoelectric conductivity from black horizon. In section 4, in the presence of magnetic field, the Hall conductivities are discussed. We apply the Sen's entropy function in our model and express the conductivities in terms of the charge and magnetic field in section 5. Finally, we conclude in section 5. 

\section{ Einstein-Maxwell-Dilaton-Axion theory in Gauss-Bonnet gravity }
We consider  the following five-dimensional action of Einstein-Maxwell-Dilaton-Axion theory in Gauss-Bonnet gravity
\bea
S=\frac{1}{2\kappa^2}\int_Md^5x \sqrt{-g}\Big(R+\la \mathcal{L}_{GB}-\frac{1}{2}(\partial{\phi})^2-V(\phi)-\frac{\Phi(\phi)}{2}\sum_{i=1}^{3}(\partial{\chi_i})^2-\frac{Z(\phi)}{4}
F_{\mu\nu}F^{\mu\nu}\Big)\label{action},
\eea
where $2\kappa^2 = 16 \pi G_5$ is the five-dimensional gravitational coupling. $\la$ is Gauss-Bonnet
coupling constant with dimension $\rm (length)^2 $ and
\bea
\mathcal{L}_{GB}=\left(R_{\mu\nu\rho\sigma}
R^{\mu\nu\rho\sigma}-4R_{\mu\nu}R^{\mu\nu}+R^2\right)
\eea
 is Gauss-Bonnet term. The linear axion fields $\chi_i(x^{\mu}) (i=1,2,3)$ are introduced to brake the  diffeomorphism symmetry, $\phi$ is dilaton field and $V$ is the potential of dilaton field.  The couplings $\Phi(\phi)$, $Z(\phi)$ and $V(\phi)$ depend on the $\phi$.  $F_{\mu\nu}=(d A)_{\mu\nu} $ is defined as  $U(1)$ gauge field strength.

The equations of motion for gravity, dilaton and Maxwell field equation are easily obtained as
\bea
&&\nabla_{\mu}\left(Z(\phi)F^{\mu\nu}\right)=0,\label{Max}\\
&&\nabla_{\mu}\nabla^{\mu}\phi-\frac{d V(\phi)}{d\phi}-\frac{1}{2}\frac{d\Phi(\phi)}{d\phi}\sum_{i=1}^{3}(\partial{\chi_i})^2-\frac{1}{4}\frac{d Z(\phi)}{d \phi}F_{\mu\nu}F^{\mu\nu}=0,\\
&&R_{\mu\nu}-\frac{g_{\mu\nu}}{2}\left(R-\frac{1}{2}(\partial{\phi})^2-V(\phi)-\frac{\Phi(\phi)}{2}\sum_{i=1}^{3}(\partial{\chi_i})^2-\frac{Z(\phi)}{4}F_{\mu\nu}F^{\mu\nu}\right),\nonumber\\
&&~~~~~~~~~~~~~-\frac{1}{2}\partial_{\mu}\phi\partial_{\nu}{\phi}-\frac{\Phi(\phi)}{2}\sum_{i=1}^{3}\partial_{\mu}{\chi_i}\partial_{\nu}{\chi_i}-\frac{Z(\phi)}{2}F_{\mu\lambda}F_\nu^{~\lambda}\nonumber\\
&&~~~~~~~~~~~~~-\la\frac{g_{\mu\nu}}{2}\mathcal{L}_{GB}+\la\left(R^2-4R_{\rho\sigma}R^{\rho\sigma}+R_{\lambda\rho\sigma\tau}R^{\lambda\rho\sigma\tau}\right)=0
\eea
The above equations of motion admits the metric  ansatz as following 
\bea
ds^2=-f(r)dt^2+\frac{1}{f(r)}dr^2+e^{2U(r)}(dx^2+dy^2+dz^2),\label{metric}
\eea
and 
\bea
\chi_i= a_ix+b_iy+c_iz ,    ~~~~~~~A=A_t(r)dt.
\eea
where the UV boundary is defined as $r\rightarrow\infty$. The gauge field has the $t$-component $A_t$ only, and we will consider the non-vanishing  magnetic field in section 4 and 5. The axion fields have only one non-zero component that is linear in the boundary coordinates for simplicity,  so the non-vanish positive constants $a_i, b_i, c_i$ take the value $a_1=b_2=c_3=k$ to admits the isotropic and homogeneous metric.  We also require the dilaton potential $V(\phi)$ satisfy the conditions
\bea
V(0)=-\frac{12}{L^2}, ~~~~V'(0)=0
\eea
in order to admit the asymptotically AdS solution.

Finally, by using the Euclidean continuation of the metric  (\ref{metric}),  the temperature of black hole is given by
\bea
T=\frac{f'(\rh)}{4\pi}.
\eea

\section{DC conductivities from horizon data}

\subsection { Electric conductivity and  thermoelectric conductivity}
Following the method \cite {Donos2014a}, to calculate the transport coefficients,  we need to generate the electric  and heat currents in the spatial directions. To achieve this, we consider   the small perturbation ansatz of the form as follows:
\bea
&&g_{tx}\rightarrow\delta g_{tx}(r) \nonumber\\
&&g_{rx}\rightarrow r^2\delta g_{rx}(r)\nonumber\\
&&A_x\rightarrow-Et+\delta A_{x}(r)\nonumber\\
&&\chi_1\rightarrow kx+\delta\chi(r).  \label{pert1}
\eea
Then linearizing the Maxwell equation, Einstein equations and Klein-Gordon equation around the metric, one can obtain the following four  equations of perturbations:
\bea
&&\delta A_{x}''+\left(\frac{f'}{f}+\frac{Z'(\phi)}{Z(\phi)}\phi'+U'\right)\delta A_{x}'+\frac{A_t'}{f}(\delta g_{tx}'+U'\delta g_{tx}+\frac{Z'(\phi)\phi'\delta g_{tx} }{Z})=0 ,\label{Max1}\\
&&\de\chi'-k \delta g_{rx}-\frac{E Z(\phi) A_t'}{k f \Phi(\phi)}=0, \label{Ein1} \\
&&\delta g_{tx}''+\frac{e^{-2U}\Phi(\phi) k^2+2f(U''-U'^2(4\la  U'f+12\la (U'^2+U'')f-3))}{f(4\la U'^2f-1)}\delta g_{tx}\nonumber\\
&&~~~~~~~~+\frac{U'\left(4\la U'(f'+U'f)+8\la U''f-1\right)}{(4\la U'^2f-1)f}\delta g_{t x}'+\frac{Z(\phi)A_t'}{4\la U'^2f-1}\delta A_{x}'=0,\label{Ein2}\\
&&\delta\chi''+\frac{f'+3U'f}{f}\delta\chi'-k\delta g_{rx}'-\frac{k(f'+3U'f)}{f}\delta g_{rx}=0.\label{KG}
\eea
Note that, from the Maxwell equations (\ref{Max}), one can define a radially conserved current as $q\equiv e^{3U}Z(\phi)A_t'$, which is the charge of the black hole. Thus linearised Maxwell equation (\ref{Max1}) results  another conserved current  as 
\bea
&&J_x=-\sqrt{-\det g}Z(\phi)F^{rx}\nonumber\\
&&~~=-e^UZ(\de g_{tx}A_t'+fA_x')\nonumber\\
&&~~=-qe^{-2U}\delta g_{tx}-e^{U}fZ(\phi)\delta A_x'
\eea
which is independent of radial coordinate $r$. So it is convenient to evaluate $J_x$  at the horizon. The equation for $\delta g_{rx}$ (\ref{Ein1}) deduces
\bea
\de g_{rx}=-\frac{E Z(\phi) A_t'}{k^2 f \Phi(\phi)}+\frac{1}{k}\de\chi'  \nonumber\\
=-\frac{q E e^{-3U}}{k^2 f \Phi(\phi)}+\frac{1}{k}\de\chi' ~ \label{grx1}
\eea

Now we have to consider the asymptotic behavior near the horizon $r = r_H$. Note that on the future horizon of the black hole, the regularity  requires the boundary condition on the horizon. Thus we
employ the ingoing Eddington-Finklestein coordinates $(u,v)$ defined as $v=t+\int  \frac{dr}{f(r)}$. For gauge field, it should be regular at the future horizon, i.e. $A_x\sim-Ev+...$ in EF coordinates.  So the perturbation field $\de A_x$ takes the asymptotic behavior near the horizon  as 
\bea
\delta A_{x}\sim-\frac{E}{4\pi T}\log(r-\rh)+\mathcal{O}(r-\rh).
\eea
Furthermore, one can see that the singular part of the metric in Eddington-Finklestein coordinates is given by 
\bea
-\frac{2\delta g_{tx}}{f(r)}drdx+2e^{2U}\delta h_{rx}drdx+2\de g_{tx} dvdx
\eea
 So from (\ref{grx1}) that $\de g_{rx}\sim1/(r-r_H)$ near horizon,  the regularity on the horizon requires that $\de g_{tx}$ behaves as 
\bea
\delta g_{tx}&\sim& e^{2U}f\delta g_{rx}|_{r\rightarrow\rh}\nonumber \\
&=&-\frac{e^{2U}(kf\Phi(\phi)\de\chi'-EZ(\phi)A_t'}{k^2 \Phi(\phi)}\bigg|_{r\rightarrow\rh}+\mathcal{O}(r-\rh)\nonumber \\
&=&\left(-\frac{qEe^{-U}}{k^2\Phi(\phi)}+\frac{e^{2U}f\chi'}{k}\right)\bigg|_{r\rightarrow\rh}.
\eea
Note that we have assumed that $\de \chi$ is analytic on the horizon.

Now we are going to calculate the value of the conserved current in order to obtain the DC conductivities. Recall the conserved current $J_x$ is constant along the radial direction $r$, one can evaluate $J_x$ at the horizon $r=r_H$. Thus the electric conductivity $\sigma$ is given by
\bea
&&\sigma=\frac{\partial J}{\partial E}\\ \nonumber
&&~~=\left(\frac{e^{-3U}q^2}{k^2\Phi(\phi)}+e^UZ(\phi)\right)\bigg|_{r\rightarrow\rh}
\eea

Next we could construct the conserved heat current to obtain the thermoelectric conductivity $\bar{\alpha}$.  One can use the $t-x$ component linearized Einstein equation and Maxwell equation to show that 
the current 
\bea
Q=e^U(f'\de g_{tx}-f\de g_{tx}')(4\la U'^2f-1)-A_t J_x
\eea
is radially constant, i.e. $\partial_r Q=0$. Thus we can evaluate the $Q$ at any where in the bulk. To see $Q$ is indeed the heat current in the $x$-direction, one can easily check that $ Q = T_{tx}-\mu J_x$, where  $T_{tx}$ is the stress tensor of the boundary. The thermoelectric conductivity is then obtained as 
\bea
&&\bar{\alpha}=\frac{\partial Q}{T\partial E}\bigg|_{r\rightarrow\rh}\\ \nonumber
&&~~=\frac{4\pi q}{k^2\Phi(\phi)}\bigg|_{r\rightarrow\rh}
\eea
Note that although the heat current $Q$ has contribution from Gauss-Bonnet term, the thermoelectric conductivity  is independent of the Gauss-Bonnet coupling $\la$, since the value of the term $\la f$ at horizon is zero.
\subsection{DC thermal and thermoelectric conductivities}
Now in order to calculate the thermoelectric and thermal conductivities,  we consider the fluctuations as follows:
\bea
&&g_{tx}\rightarrow \delta g_{tx}(r)+t\delta h(r) \nonumber\\
&&g_{rx}\rightarrow e^{2 V(r)}\delta g_{rx}(r)\nonumber\\
&&A_x\rightarrow\delta A_{x}(r)+ t\delta a(r)\nonumber\\
&&\chi_1\rightarrow kx+\delta\chi(r).  \label{pert2}
\eea
Similarly, the linearised Maxwell equation involves a conserved current
\bea
\tilde{J}_x=-e^{U}Z(\phi)(t\de h+\de g_{tx})A_t'+f(t\de a'+\de A_x'),\label{curr3}
\eea
and the  $r-x$ component of the linearised Einstein equation  gives
\bea
-\frac{1}{2}k^2\Phi(\phi)\de g_{rx}+\frac{1}{2f}(Z(\phi) A_t' \de a+\de h'- 2U'\de h)-2\la U'^2(\de h'-2U'\de h)+\frac{1}{2}k\Phi(\phi)\chi'=0
\eea

As last section, we want to construct the conserved heat current $Q$. To achieve this, we linearize  the $t-x$ component of the Einstein equation and combine with the linearized Maxwell equation, 
the  conserved current could be defined as
\bea
\tilde{Q}=e^U\left[f'\de g_{tx}+t  f'\de h -f(\de g_{tx}'+t \de h')\right](4\la f U'^2-1)-A_t \tilde{J}_x
\eea
 We choose the perturbations take the form as  $\de h(r)=-\gamma f(r)$, $\de a(r)=-E+ \gamma A_t(r)$ to cancel the time-dependent part of those two conserved currents. Then one can get the
  conserved  electric current as
  \bea
  \tilde{J}_x=-e^UZ(\phi)(A_t'\de g_{tx}+f\de A_x') \label{new J}
  \eea
  Since $\tilde{J}_x$ is constant along $r$ direction,  we can see that $\tilde{J}_x$ is the electric current in $x$ direction by evaluating at boundary $r=\infty$.
  The heat current is  given by
  \bea
  \tilde{Q}=(4\la f U'^2-1)(f'\de g_{tx}-f\de g_{tx}')-A_t \tilde{J}_x \label{new Q}
  \eea
As explained before, $\tilde{Q}$ is the time-independent part of the heat current as well.
  
Before evaluate the electric current and heat current on the horizon, we need to know  the asymptotic behaviours of the perturbations near the horizon. We switch the metric in the Kruskal coordinates $(\tilde{U},\tilde{V})$. The singular part of the metric is  
\bea
&&\frac{2}{f'(\rh)}\left(-e^{-\frac{f'(\rh)}{2}+\frac{f'(\rh)}{2}(u+v)+2\tilde{U}}f\de g_{rx}+e^{-\frac{f'(\rh)}{2}v+\frac{f'(\rh)}{2}(u+v)}\de g_{tx}+e^{-\frac{f'(\rh)}{2}+\frac{f'(\rh)}{2}(u+v)}t\de h \right)d\tilde{U}dx \nonumber\\
&&~~~~~~~~~~~~~~~+\frac{2}{f'(\rh)}\left(e^{\frac{-f'(\rh)}{2}+\tilde{U}}f\de g_{rx}+e^{\frac{-f'(\rh)}{2}v}\de g_{tx}+e^{\frac{-f'(\rh)}{2}v}t\de h\right)d\tilde{V}dx 
\eea
So the regularity on the black horizon requires that  
\bea
\de A_x \sim -\frac{E}{4\pi T}\ln(r-\rh),\\
\de g_{tx}\sim e^{2U}f\de g_{rx} |_{r\rightarrow\rh}-\frac{\gamma f}{4\pi T}\ln(r-\rh)
\eea

Have obtained the behavior of perturbation  near horizon, we can evaluate the electric current (\ref{new J}) and heat current (\ref{new Q}) on the horizon and get the DC conducivities as
 \bea
&& \alpha=\frac{1}{T}\frac{\partial \tilde{J}_x}{\partial \gamma}=\frac{4\pi q}{k^2 \Phi(\phi)}\big|_{r\rightarrow\rh},\nonumber\\
 &&\bar{\kappa}=\frac{1}{T}\frac{\partial \tilde{Q}}{\partial \gamma}=\frac{16\pi^2Te^{3U}}{k^2\Phi(\phi)}\big|_{r\rightarrow\rh}.
 \eea
One can also find that thermal conductivity and thermoelectric conductivity are the Gauss-Bonnet coupling independent.
 \section{Hall conductivity from horizon data} 
 In this section, to derive general analytic expressions for  Hall conductivity, we shall to consider  our holographic model in the presence of a magnetic field. So the Maxwell field takes the form as
 \bea
 A=A_t(r)dt+Bxdy.
 \eea
  One could  consider   the small perturbation ansatz around the black hole 
 \bea
 &&g_{tx}\rightarrow e^{2V}\delta g_{tx}(r),~~~~~~~~~~~
 g_{ty}\rightarrow e^{2V}\delta g_{ty}(r)\nonumber\\
&&g_{rx}\rightarrow e^{2 V(r)}\delta g_{rx}(r),~~~~~~~
g_{ry}\rightarrow e^{2 V(r)}\delta g_{ry}(r)\nonumber\\
&&A_x\rightarrow\delta A_{x}(r)-Et,~~~~~~~
A_y\rightarrow\delta A_{y}(r)+B x\nonumber\\
&&\chi_1\rightarrow kx+\delta\chi(r).  \label{pert2}
\eea
By using the Maxwell equations, one can define the conserved currents as
\bea
J_x=-e^UZ\left( e^{2U}A_t'\de h_{tx}+fB\de h_{ry}+f\de A_x'\right),\nonumber\\
J_y=-e^UZ\left( e^{2U}A_t'\de h_{ty}-fB\de h_{rx}+f\de A_y'\right). \label{n current}
\eea
Both of these two currents are independent of the radial coordinate, so we will evaluate them on the horizon. 

To proceed the evaluation, one can following the same discussion as before  to   analysis the asymptotic behavior for the those perturbation fields near horizon. The  regularity on the horizon require that 
\bea
&&\delta A_{x}\sim-\frac{E}{4\pi T}\log(r-\rh)+\mathcal{O}(r-\rh),\nonumber\\
&&\de h_{tx}\sim f \de h_{rx}+\mathcal{O}(r-\rh), ~~~~\de h_{ty}\sim f \de h_{ry}+\mathcal{O}(r-\rh) \label{bdy3}
\eea
and the $\de \chi$ is analytic on the horizon.

On the otherhand, one can combine the  $r-x$, $x-x$ components of linearised  Einstein equations 
\bea
\frac{-E+B\de h_{ty}}{f}ZA_t'+Be^{-2U}Z\left(\de A_y'-B\de h_{rx}\right)+k\Phi(\de\chi'-k\de h_{rx})=0 \label{EQ1}
\eea
Similarly,  one can get 
\bea
\frac{B\de h_{tx}}{f}ZA_t'+k^2\Phi\de h_{ry}+Be^{-2U}Z(B\de h_{ry}+\de A_x')=0 \label{EQ2}
\eea
from the for the $r-y$, $y-y$ components of Einstein equation. Note that above equations have no contribution from Gauss-Bonnet term. Plugging the boundary condition (\ref{bdy3}) into (\ref{EQ1}-\ref{EQ1}), one can obtain two equations about $\de h_{tx}$, $\de h_{ty}$ on the horizon
\bea
(B^2Z+e^{2U}k^2\Phi)\de h_{tx}+e^{2U}\left(E-B\de h_{ty}\right)ZA_t'\bigg|_{r\rightarrow\rh}=0\nonumber\\
(e^{2U}k^2\Phi+B^2Z)\de h_{ty}+B Z(-E+e^{2U}\de h_{tx}A_t')\bigg|_{r\rightarrow\rh}=0
\eea
which could give the $\de h_{tx}$, $\de h_{ty}$ on the horizon.

Finally, put those above results  back to the electrical currents expressions (\ref{n current}) and evaluate them on the horizon, we can get the Hall conductivities as
\bea
&&\sigma_{xx}=\frac{J_x}{E_x}=\frac{e^{3U}k^2\Phi (\rho^2+e^{2U}B^2Z^2+e^{4U}k^2Z\Phi)}{B^2\rho^2+e^{2U}(B^2Z+e^{2U}k^2\Phi)^2},\nonumber\\
&&\sigma_{xx}=\frac{J_y}{E_x}=\frac{B\rho  (\rho^2+e^{2U}B^2Z^2+2e^{4U}k^2Z\Phi)}{B^2\rho^2+e^{2U}(B^2Z+e^{2U}k^2\Phi)^2}. \label{hall}
\eea
where we have used the boundary charge density $\rho=-Z e^{3U}A_t'$. So one can easily see that Gauss-Bonnet correction does not contribute to the Hall conductivities as well. Note that the explicit expression for conductivities will not be determined  until we find the  exact solution for black hole.

\section{DC conductivities and Entropy function} 
 In this part, we will use the Sen's entropy function formalism \cite{Sen1} to calculate the conductivities from the information near black hole horizon. By solving the attractor equations, which are derived  from the entropy function, one can get the values of the matter fields on the horizon of the black hole without knowing the full solution of the metric. This approach  applied in holography  was proposed firstly in \cite{Erd}, and subsequently in \cite{Nas1,Nas2}, in which the Maxwell-Dilaton theory  and Topological Maxwell theory were discussed respectively. Both of these models were in the framework of classical
 Einstein gravity. Here as explained in the introduction, we will consider the holographic model in Gauss-Bonnet gravity to see the effect of stringy correction in the semi-classical limit.
 
 \subsection{Sen's entropy function formalism}
 As discussion in the seminal paper \cite{Erd}, one could consider a class of extremal black hole, which have geometry of $AdS_2 \times S^{D-2}$ in the near horizon limit. One can define the entropy function
 via the Legendre transformation of the integral for the Lagrangian density evaluated for this black hole background over the sphere. Note that, as pointed out in \cite{Erd}, although the original discussion in \cite{Sen1} relies on the geometry of  $AdS_2 \times S^{D-2}$, it can be valid for the planar AdS black hole horizon with geometry $AdS_2 \times R^{D-2}$. 
 
 Let's follow  the entropy function formalism method  and apply it in five dimensional Gauss-Bonnet gravity here.  Consider the extremal black hole solution near horizon takes the form 
 \bea
 ds^2=v\left(-r^2dt^2+\frac{dr^2}{r^2}\right)+w(dx^2+dy^2+ dz^2)
 \eea
 where $v$ is the  $AdS_2$ radius and $w$ is $R^3$ radius. The attractor mechanism shows that the value of the matter  fields  on the horizon are independent of the values at boundary. We  denote the scalar fields and Maxwell field take the values on the horizon as $\phi_s=u_s$, $F_{rt}=e$, $F_{xy}=B$. We will see that these constants are  completely fixed by the charge of the black hole and magnetic field.
 
 First we define a functional from the gravitational Lagrangian density as $f(u_s,v,w,e)=\int dxdydz \sqrt{-g}\cal{L}$, where the integral is carried out on the horizon. So for our action (\ref{action}), the functional $f(u_s,v,w,e)$ is obtained as
 \bea
 f(u_s,v,w,e)=-\frac{\textrm{Vol} \mathbb{R}^3}{4\kappa^2 v\sqrt{w}}\left(2v^2 w^2V(u_s)+Z(u_s)(B^2v^2-e^2w^2)+4vw^2+3k^2v^2w\Phi(u_s)\right)
 \eea
 where we denote $\textrm{Vol} \mathbb{R}^3$ as the volume of $\mathbb{R}^3$, and $u_s$ is the value of the dilaton field on the horizon. It should be pointed out that, by contrast with the results in
\cite{Morales}, there is no Gauss-Bonnet term contributions here since we consider the geometry $AdS_2 \times R^{D-2}$  instead of $AdS_2 \times S^{D-2}$.
 
 Thus  by using Legendre transformation, one can define the entropy function as
 \bea
 F(u_s, v, w, Q_e)=2\pi (e Q_e -  f(u_s, v, w, e)) \label{entropy}
 \eea
 with charge $Q_e$ defined as 
 \bea
 &&Q_e=\frac{\partial f(u_s, v, w, e)}{\partial e}\nonumber\\
 &&~~~~=\frac{ w^{3/2} e Z(u_s)}{2\kappa^2v}\textrm{Vol} \mathbb{R}^3. \label{charge}
 \eea
 Besides, the Einstein equations and the equations for scalar field  near horizon is given by 
 \bea
\frac{\partial f(u_s,v,w,e)}{\partial v }=0, ~~\frac{\partial f(u_s,v,w,e)}{\partial u_s }=0
 \eea
 
 Now we can get attractor equations  by extremizing the entropy function with respect to $u,v,w$ as
 \bea
 \frac{\partial F(u_s, v, w, Q_e)}{\partial u_s}=0, ~~\frac{\partial F(u_s, v, w, Q_e)}{\partial v}=0,~~ \frac{\partial F(u_s, v, w, Q_e)}{\partial w}=0,
 \eea
 which give three algebraical  attractor equations as
\bea
&&wV(u_s)+\frac{e^2 w Z(u_s)}{2v^2}+\frac{B^2}{2w}Z(u_s)+\frac{3\Phi(u_s)k^2}{2}=0,\label{eq1}\\
&&w+\frac{vw}{2}V(u_s)-\frac{e^2 w }{4v}Z(u_s)-\frac{B^2v}{12w}Z(u_s)+\frac{k^2v}{4}\Phi(u_s)=0,\label{eq2}\\
&&vw^2\frac{dV(u_s)}{du_s}+\frac{B^2v}{2}\frac{dZ(u_s)}{du_s}-\frac{e^2w^2}{2v}\frac{dZ(u_s)}{du_s}+\frac{3k^2vw}{2}\frac{d\Phi(u_s)}{du_s}=0\label{eq3}.
 \eea
 and the near horizon electric field $e$ from entropy function as 
  \bea 
  e=\frac{1}{2\pi} \frac{\partial F}{\partial Q_e}
  =\frac{2\kappa ^2Q_ev}{w^{3/2}Z(u_s)\textrm{Vol} \mathbb{R}^3}
  \eea
  Combine the  attractor equations (\ref{eq1}), (\ref{eq2}),  one can obtain the 
  \bea
  &&Z(u_s)=\frac{3v w(2w-k^2\Phi(u_s)v)}{2B^2v^2+3e^2w^2},\nonumber\\
  &&V(u_s)=-\frac{6w(B^2v^2+e^2w^2)+3k^2\Phi(u_s)v(B^2v^2+2e^2w^2)}{4B^2v^3w+6e^2vw^3},
  \eea
 Bring them back to the entropy function (\ref{entropy}) and use the expression for $Q$ (\ref{charge}), we can get the entropy function as
  \bea
  s=\frac{2\pi w^{3/2}}{\kappa^2}\textrm{Vol} \mathbb{R}^3=\frac{A}{4G}
  \eea
  where $A=w^{3/2}\textrm{Vol} \mathbb{R}^3$ is the area of the horizon, so it matches the Bekenstein-Hawking formula.  Note that unlike the results in \cite{Morales}, our result has no GB term contributions in the entropy due to the planar horizon,  which is consistent with the result in \cite{cai}.

As mentioned before, the attractor mechanism implies that the values of the fields on the horizon are  dependent  only on the charge of the black hole and magnetic field. So by using the attractor equations, one can get the conductivities at zero temperature in terms of the charge density and magnetic field as well. For simplicity, we take $\kappa=1,$ $\Phi(\phi)=0$ and consider the potential $V(\phi)=-12/L^2$ as an example.
 In this case, the solutions for attractor equations are given by
 \bea
 Z(u_s)=2\left(\frac{6q}{B^6L^2}\right)^{1/5}, w=\left(\frac{Bq}{6L^2}\right)^{2/5}, e=\frac{1}{5}\left(\frac{9B^3L^6}{8q^2}\right)^{1/5}, v=\frac{L^2}{10}.
 \eea   
 where $q=Q/\textrm{Vol} \mathbb{R}^3$ could be identified as the charge density $\rho$. Note that  all the fields on the horizon depend only on   the charge of the black hole and magnetic field. So we can get explicit form for the Hall conductivities  (\ref{hall}) at zero temperature 
 \bea
 \sigma_{xx}=0,~~~~  \sigma_{xy}=\frac{\rho}{B},
 \eea
 even without knowing the explicit solutions of the extremal black hole.

\section{Summary}  
  In this paper, by using AdS/CFT,  we studied the  DC  conductivities and Hall conductivities for Einstein- Maxwell-Dilaton in Gauss-Bonnet gravity  with momentum relaxation. We  derived the analytic expressions for the DC electric conductivity, thermal and thermoelectric conductivities  and Hall conductivities.   So we believe that for the  holographic gravity model with Gauss-Bonnet term, one  can still get the DC conductivities from horizon data. Beside that, we also show that the DC  conductivities and Hall conductivities are not dependent on the Gauss-Bonnet coupling. Finally, we use the Sen's entropy function formalism to derive the  conductivities  as a function of the charges of the black hole and magnetic field. 
  
  There are still some interesting work for future investigation. For example, here we consider the  Gauss-Bonnet gravity coupled with a nontrivial dilaton field, and get the  Gauss-Bonnet coupling independent conductivities. One can ask when we consider the Gauss-Bonnet gravity with a non-minimally coupled dilaton \cite{cai2}, it is reasonable to believe that the conductivities should be dependent on the Gauss-Bonnet coupling and examine how the above results are modified with stringy corrections. 
  
\section*{Acknowledgments}
 This work was  supported by NSFC, China (No.11905185).

\end{document}